\documentclass[aps,prb,preprint,superscriptaddress]{revtex4}

\usepackage{graphicx}
\usepackage{float}
\usepackage[caption=false,subrefformat=parens,labelformat=parens]{subfig}

\begin{document}
\title{Unconventional superconductivity and an ambient-pressure magnetic quantum critical point in single-crystal LaNiC$_2$}
\author{J. F. Landaeta}
\author{D. Subero}
\author{P. Machado}
\affiliation{Centro de F\'{\i}sica, Instituto Venezolano de
	Investigaciones Cient\'{\i}ficas, Apartado 20632, Caracas
	1020-A, Venezuela}

\author{F. Honda}
\affiliation{Institute for Materials Research, Tohoku University, Oarai, Ibaraki 311-1313, Japan}

\author{I. Bonalde}

\affiliation{Centro de F\'{\i}sica, Instituto Venezolano de
	Investigaciones Cient\'{\i}ficas, Apartado 20632, Caracas
	1020-A, Venezuela}

\date{\today}
\begin{abstract}
Superconductivity in noncentrosymmetric LaNiC$_2$ is expected to be induced by electron–-phonon interactions due to its lack of magnetic instabilities. The non-Bardeen-Cooper-Schrieffer (BCS) behaviors found in this material call into question the long-standing idea that relates unconventional superconductivity with magnetic interactions. Here we report magnetic penetration-depth measurements in a high-purity single crystal of LaNiC$_2$ at pressures up to 2.5 GPa and temperatures down to 0.04 K. At ambient pressure and below 0.5$T_c$ the penetration depth goes as $T^4$ for the in-plane and $T^2$ for the out-of-plane component, firmly implying the existence of point nodes in the energy gap and the unconventional character of this superconductor. The present study also provides first evidence of magnetism in LaNiC$_2$ by unraveling a pressure-induced antiferromagnetic phase inside the superconducting state at temperatures below 0.5 K, with a quantum critical point around ambient pressure. The results presented here maintain a solid base for the notion that unconventional superconductivity only arises near magnetic order or fluctuations.
\end{abstract}


\maketitle

\section{Introduction}

In conventional BCS superconductors magnetism appears as a competing phenomenon. However, experiments show that in almost all known bulk unconventional superconductors, such as Sr$_2$RuO$_4$,\cite{Sidis1999} high-$T_c$ cuprates,\cite{Keimer2015} iron pnictides,\cite{Si2016} CePt$_3$Si,\cite{Bauer2004} UPt$_3$,\cite{Aeppli1988} and other heavy fermions,\cite{Pfleiderer2009} magnetism coexists with or is in close proximity to the superconducting phase. This group excludes the nonmagnetic LaNiC$_2$ and PrOs$_4$Sb$_{12}$, although in the latter compound neutron-diffraction data revealed a small field-induced antiferromagnetic ordering.\cite{Kohgi2003} Since unconventional superconductivity is defined as the breaking of additional symmetries besides the gauge symmetry U(1), Li$_2$Pt$_3$B \cite{Yuan2006} is not considered here. These experimental findings have led to the long-standing idea that unconventional superconductivity should be tightly connected with magnetic instabilities. In this sense the possible existence of non-BCS superconductivity in LaNiC$_2$,\cite{Lee1996,Bonalde2011,Hillier2009} which has been shown to be in a nonmagnetic ground state at ambient pressure and zero field, \cite{Iwamoto1998,Pecharsky1998} introduces a puzzling and highly fundamental issue: Can unconventional superconductivity exist in the complete absence of a magnetic environment?

The lack of magnetism in LaNiC$_2$ (as well as in YNiC$_2$) is surprising, since in most Ni-based compounds the $3d$-Ni magnetic moment yields magnetic ordering. However, in LaNiC$_2$ the Ni atom does not have a moment.\cite{Kotsanidis1989,Schafer1992} Magnetism can be sometimes stabilized by doping but this does not occur in LaNiC$_2$ when doped with Y, Th, or Cu.\cite{Liao2009,Sung2008,Lee1997} On the other hand, unconventional responses have been reported for LaNiC$_2$ in low-temperature specific heat \cite{Lee1996} and magnetic penetration-depth \cite{Bonalde2011} studies, and supported by evidence of time reversal symmetry (TRS) breaking.\cite{Hillier2009} However, there is no consensus about these results in the superconducting phase. Measurements using nuclear-quadrupole-resonance (NQR) 1/T$_1$ \cite{Iwamoto1998} suggested that LaNiC$_2$ is a conventional BCS superconductor. It was shown that magnetic impurity substitutions as low as 3 \% weaken superconductivity,\cite{Prathiba2016,Katano2017} which also supports a conventional behavior. Previous studies were mostly performed on different quality polycrystalline samples that were shown to influence the superconducting properties.\cite{Bonalde2011} The recent growth of single crystals \cite{Hirose2012} opens up new opportunities to elucidate the above-stated issue. Heat-capacity measurements \cite{Hirose2012} in single crystals appear to corroborate conventional superconductivity in this compound.

By measuring the magnetic penetration depth -widely accepted as one of the most powerful probes to determine the energy gap structure and the unconventional nature of superconductors- of a high-purity single crystal down to 40 mK and up to 2.48 GPa, here we provide a very strong evidence of unconventional superconductivity and the first indication of a pressure-induced magnetic phase in LaNiC$_2$. We did not observe any magnetic signal at ambient pressure, confirming previous results. Our observations point to the existence of a quantum critical point (QCP) around the pressure $p=0$ and uphold the very important thought that unconventional responses are only displayed in the vicinity of magnetic instabilities.

\section{Methodology}

LaNiC$_2$ crystallizes in the CeNiC$_2$-type orthorhombic structure (space group $Amm2$) with La plane at $x=0$ and NiC$_2$ plane at $x=0.5$ sequentially stacked along the $a$-axis.\cite{Hirose2012} The structure lacks inversion symmetry along the $c$-axis. The single crystal was grown by the Czochralski method in a tetra-arc furnace and grew in a cylindrical ingot of about 2 mm in diameter. The purity of the starting materials was 3N for La, 5N5 for C and 4N8 for Ni. A sample was cut to dimensions $0.5\times 0.5\times 0.3$ mm$^{3}$, then polished and used in all measurements. Other crystalline pieces were measured to corroborate the results.

At ambient pressure the penetration depth was measured with a 13 MHz tunnel diode oscillator system.\cite{Bonalde2005} High-resolution magnetic penetration depth measurements under pressure were carried out in a newly developed system \cite{Landaeta2017} which utilizes a 13 MHz tunnel diode oscillator coupled to a self-clamped hybrid-piston-cylinder cell made of nonmagnetic BeCu and NiCrAl alloys. Both experimental setups were extensively checked and compared with each other to make sure they yield identical results. To within a calibration factor, the oscillator frequency shift with respect to the value at the lowest temperature is proportional to the penetration depth deviation, $\Delta f(T) \propto \Delta\lambda(T)$. For the configuration $H\parallel a$ axis we probe the in-plane magnetic penetration depth $\Delta\lambda_\parallel(T)$, whereas for $H\perp a$ axis we probe the out-of-plane magnetic penetration depth $\Delta\lambda_\perp(T)$.

Fully relativistic first-principles electronic structure calculations were carried out using a full potential linearized augmented plane-wave (FPL-APW) method in the framework of density functional theory as implemented in the WIEN2k package.\cite{Blaha2014}

\section{Ambient pressure results}

\begin{figure}[t]
	\centering
	\scalebox{0.6}{\includegraphics{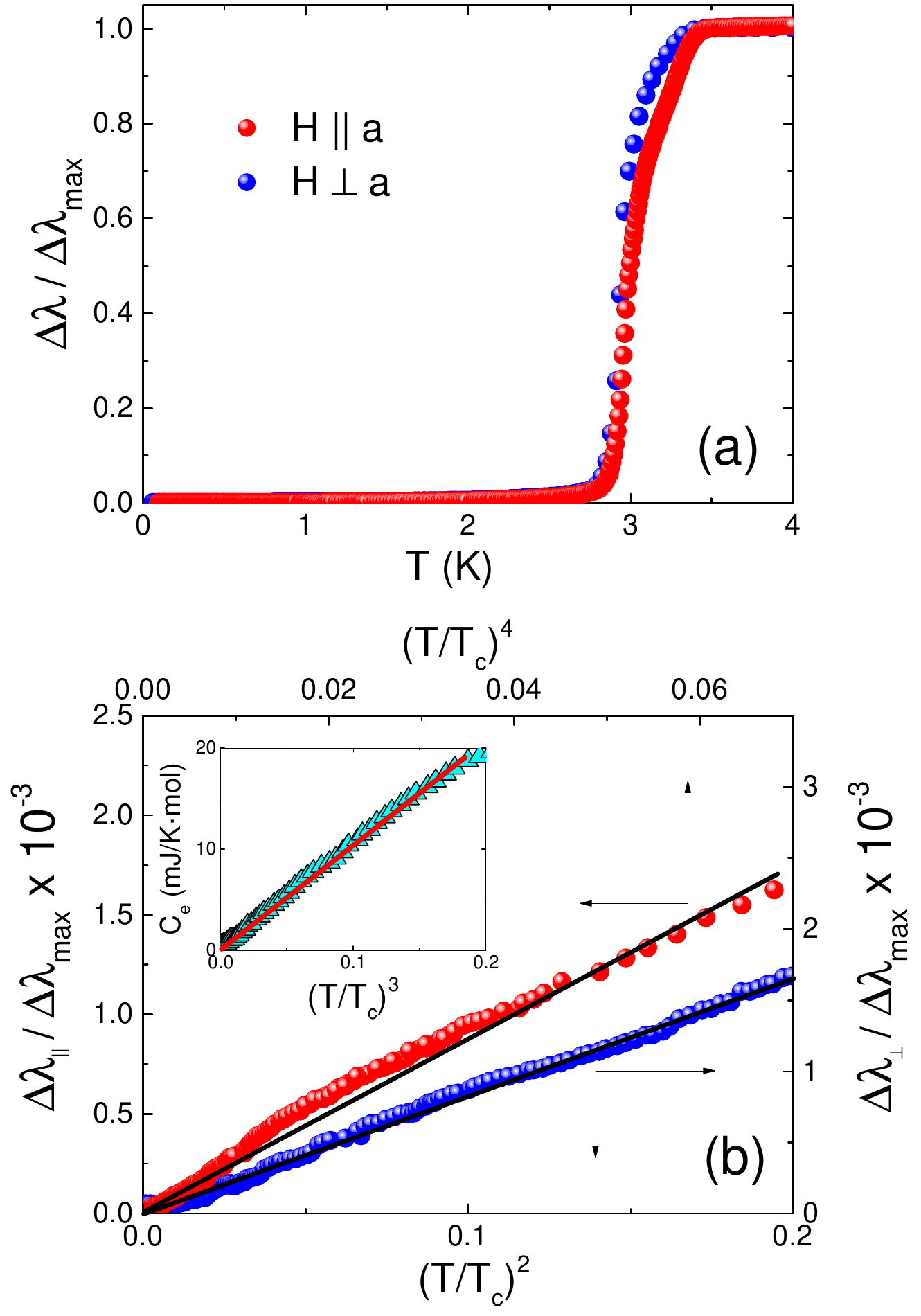}}
	\caption{\label{fig:Lambda}{(Color online) Penetration depth shift as a function of temperature in LaNiC$_2$ single crystals at ambient pressure. (a) In-plane ($H\parallel a$ axis) and (b) out-of-plane ($H\perp a$ axis) magnetic penetration depth. $\Delta\lambda_{max}$ is the total penetration-depth shift from $T_c$ down to the lowest temperature. (b) In-plane and out-of-plane penetration depth in the low-temperature region below $0.5T_c$ (and assuming $T_c=3$ K), from where it is evident that $\Delta\lambda_{\parallel} \propto (T/T_c)^4$ and $\Delta\lambda_{\perp} \propto (T/T_c)^2$.  The inset shows that the digitized specific-heat data of Hirose \textit{et al.} \cite{Hirose2012} can also be fitted to a power law $\propto (T/T_c)^3$ in the low-temperature limit. Both penetration-depth and heat-capacity measurements in single crystals provide evidence of point nodes in the energy gap of LaNiC$_2$. Lines are guides to the eye.}}
\end{figure}

The normalized penetration-depth deviations $\Delta\lambda(T)$ for magnetic fields $H\parallel a$ and $H\perp a$ are shown in Fig.~\ref{fig:Lambda}(a).  It is observed that for both field directions the diamagnetic onset occurs around 3.4 K. This is consistent with the transition onsets in magnetic susceptibility and electrical resistivity of poly- and single-crystal samples.\cite{Lee1996,Pecharsky1998,Hirose2012,Chen2013,Katano2014} Heat-capacity measurements indicate that bulk superconductivity sets in around 2.7 K, with a transition width of 0.5 K.\cite{Lee1996,Pecharsky1998,Hirose2012} Since our diamagnetic transitions are quite broad we take their midpoint around 3 K as the $T_c$.

Figure \ref{fig:Lambda}(b) displays the low-temperature regime below 1.5 K and down to 0.04 K, where it is seen that the in-plane penetration depth $\Delta\lambda_\parallel(T) \propto (T/T_c)^4$ and the out-of-plane penetration depth $\Delta\lambda_\perp(T) \propto (T/T_c)^2$. In orthorhombic crystals having a finite spin-orbit coupling, these power laws are expected for superconducting energy gaps with point nodes at the poles ([100] axis).\cite{Annett2011,Mukherjee2014} Our finding is consistent with previous heat-capacity and penetration-depth results in polycrystalline samples \cite{Lee1996,Bonalde2011} which suggested the presence of nodes in the energy gap. We reanalyzed the recent heat-capacity data obtained in single crystals \cite{Hirose2012} and found that they nicely follow a $(T/T_c)^3$ power law (inset of Fig.~\ref{fig:Lambda}(b)), which is indicative of point nodes.

To further analyze the penetration depth results, we performed numerical simulations of the normalized superfluid density $\rho=\lambda^2(0)/\lambda^2(T)$ in the local approximation

\begin{equation}
\rho_{ij}=\frac{n^s_{ij}}{n}=3\left\langle\hat{k_i}\cdot\hat{k_j}\left[1-\int d\xi\left(-\frac{df}{dE_\mathbf{k}}\right)\right]\right\rangle_{\hat{k}}.
\end{equation}

\begin{figure}
	\centering
	\scalebox{0.36}{\includegraphics{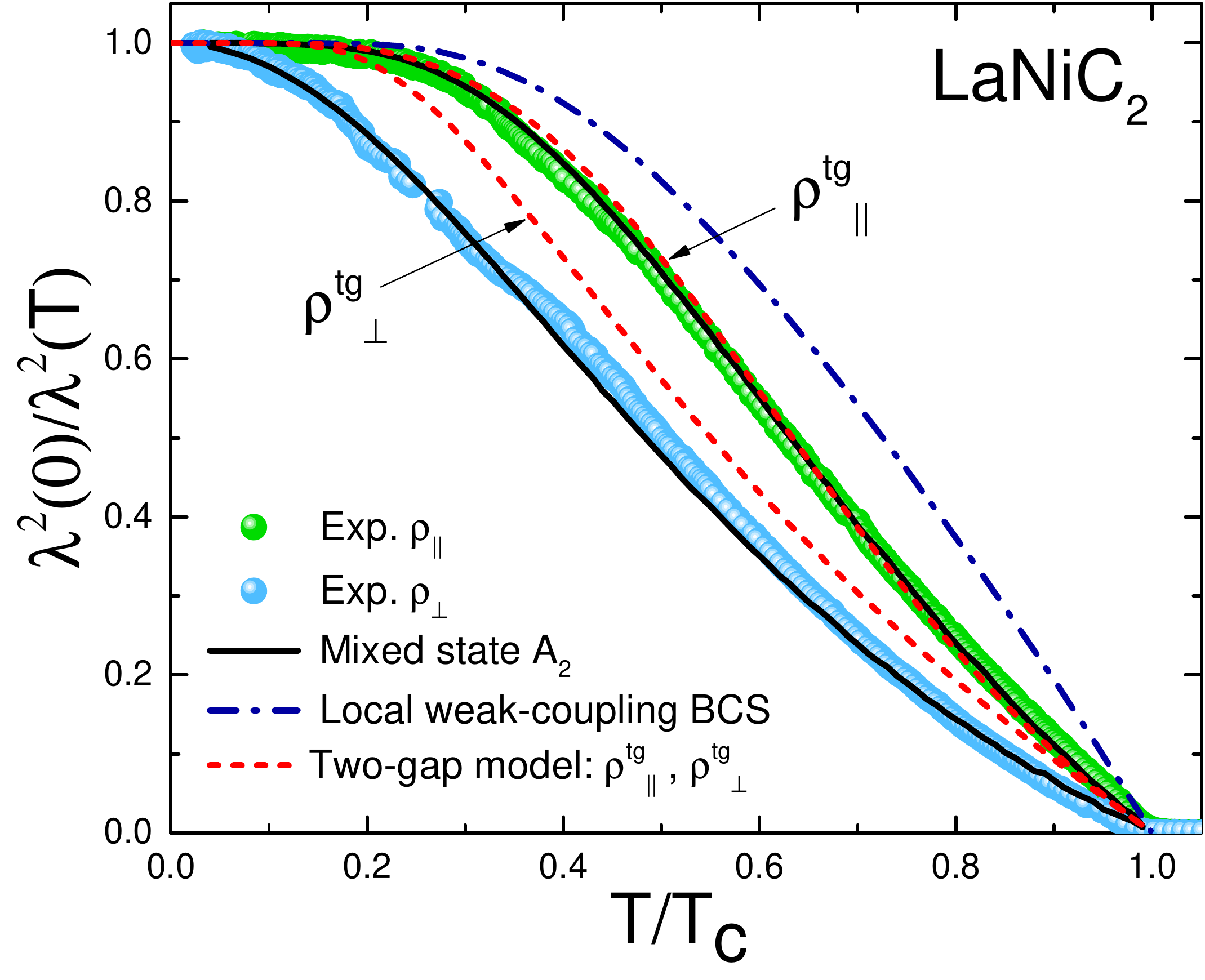}}
	\caption{\label{fig: SuperFluid}{(Color online) Superfluid density of LaNiC$_2$ compared with mixed state $A_2$, two-gap, and BCS models in the parallel and perpendicular orientations of the magnetic field with respect to the $a$-axis. We assumed for this graph $T_c=3$ K. The agreement of the experimental data with the state $A_2$ model is outstanding and gives support for the existence of point nodes in the energy gap of LaNiC$_2$.}}
\end{figure}

\noindent Here $\langle\dots\rangle$ represents an angular average over the Fermi surface (with spherical symmetry as simplified model), $f$ is the Fermi function,  $\hat{\mathbf{k}}$ is the momentum, and $E(\mathbf{k})$ is the total energy. The superfluid density of LaNiC$_2$ was obtained by estimating $\lambda(0)\approx$ 1230 $ \mathring{A}$ from $\gamma_n\approx7 \mbox{ mJmol}^{-1}\mbox{K}^2$ and Hc$_2$(0)$\approx$ 1250 Oe.\cite{Lee1996,Iwamoto1998} Since LaNiC$_2$ is an isotropic superconductor,\cite{Hirose2012} we used the same value of $\lambda(0)$ for both field orientations. In Fig.~\ref{fig: SuperFluid} we compare the normalized superfluid density of LaNiC$_2$ with numerical simulations of the mixed state $A_2$ model (with the $z$-axis parallel to the crystallographic $a$-axis) of the point group $C_{2\nu}$ with spin-orbit coupling (SOC) included. This state is an admixture of spin-singlet and spin-triplet states and has symmetry-required point nodes along the $z$-axis.\cite{Mukherjee2014} A two-gap model and the local weak-coupling BCS approximation are also included in Fig.~\ref{fig: SuperFluid} for comparison. The $A_2$ model compares remarkably well with both components (in-plane and out-of-plane) of the superfluid density. A two-gap approach fails to fit the perpendicular component for any set of parameters of the density-of-states contributions. We note that the pure $p$-wave triplet component of the state $A_2$ (also with point nodes along [100]) fits equally well with the experimental data (with slightly different superconducting parameters), making it difficult to distinguish between the pure triplet state and the mixed state. In any case, the spin-triplet state seems to be dominant and the results clearly indicate that LaNiC$_2$ has point nodes along the crystallographic $a$-axis. Thus both the penetration depth in the true low-temperature limit and the superfluid density in the entire temperature region provide strong evidence of point nodes at the poles (along the crystallographic $a$-axis) of the superconducting energy gap, firmly establishing that LaNiC$_2$ is an unconventional superconductor. The presence of point nodes poses a striking problem in regard to TRS breaking in this system, since broken-TRS states occur in nonunitary triplet channels with gap functions having line nodes and gapless excitations.\cite{Hillier2009,Quintanilla2010,Mukherjee2014} This inconsistency adds to the current debate on the violation of TRS in an orthorhombic crystal with finite SOC.\cite{Hillier2009,Quintanilla2010,Mukherjee2014}

\section{Finite pressure results}

\begin{figure}
	\scalebox{0.6}{\includegraphics{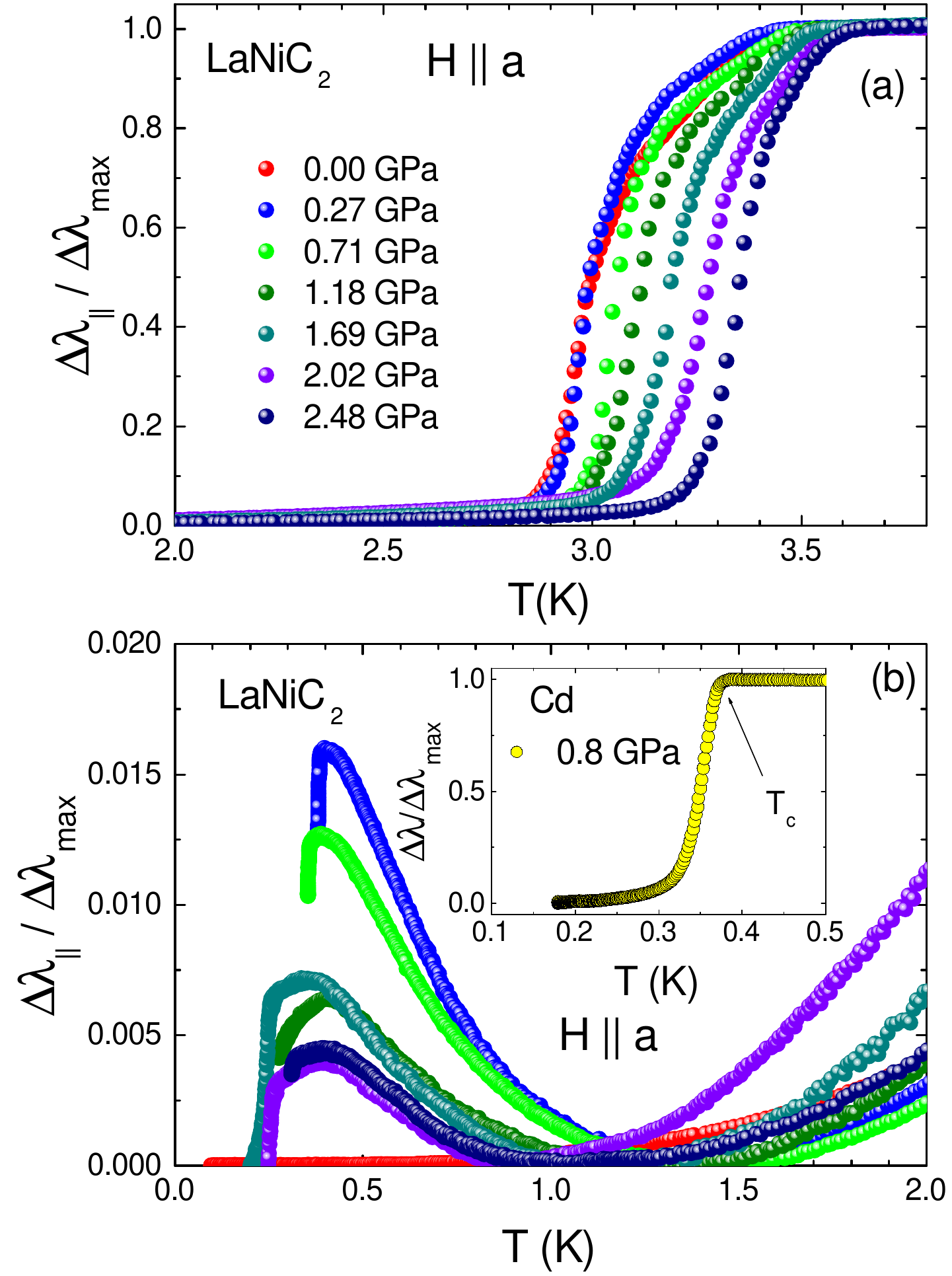}}		
	\caption{\label{fig:Pressure Sample A-1}{(Color online) Penetration depth as a function of temperature and pressure in LaNiC$_2$. (a)  In-plane response for various pressures up to 2.48 GPa. With pressure the superconducting transition seems to become sharper and the critical temperature slightly increases. (b) Low-temperature in-plane penetration depth showing sudden upturns followed by abrupt drops at the lowest temperatures. These behaviors are clear indications of the appearance of a pressure-induced magnetic order in LaNiC$_2$. The inset show for comparison cadmium data taken in the same experimental system.}}
\end{figure}

In order to study the pressure response of the superconducting phase we measured the penetration depth in the configuration $H\parallel a$ up to 2.48 GPa. Figure~\ref{fig:Pressure Sample A-1}(a) exhibits the results for the high-temperature region. From this figure we see that the critical temperature goes up from 3 to 3.35 K and the superconducting transition becomes sharper as the applied pressure increases. The rise in $T_c$ is in complete agreement with earlier resistivity measurements.\cite{Katano2014} Since LaNiC$_2$ was thought to be a rather simple alloy this result at first was unanticipated. In simple metals and their alloys, pressure causes a decrease of the density of states at the Fermi level $N(E_F)$ and so diminishes $T_c$ according to the McMillan equation $k_BT_c=(\hbar\omega_D/1.45)\exp{[-1.04(1+\Lambda)/(\Lambda-\mu^\star(1+0.62\Lambda))]}$. Here, $\Lambda = N(E_F)<\alpha^2>/M<\omega^2>$,  $\omega_D$ is the Debye frequency, $\Lambda$ is the electron-phonon coupling constant, $\mu^\star$ is the Coulomb pseudopotential, $\alpha$ is the electron-phonon matrix element, $M$ is the atomic mass, and $\omega$ is the phonon frequency. Recent first-principles band-structure calculations under increasing external pressure in LaNiC$_2$ yielded a decrease in the density of states but an increase in $\Lambda$ and, therefore, in $T_c$.\cite{Wiendlocha2016} The numerical data, however, markedly fail to reproduce the entire $T-P$ phase diagram, calling into question the pure electron-phonon interaction as the driving mechanism of superconductivity in this material.

On the other hand, the sharpening of the superconducting transition with pressure may discard sample inhomogeneities as the origin of the broad transition at ambient pressure. This pressure-induced effect then opens the question about the physical source of the broadening. We note that a similar behavior of the pressure dependence near $T_c$ was observed in penetration-depth studies on CeIrSi$_3$ and CeRhSi$_3$.\cite{Landaeta2017} In these Ce-based compounds the transition becomes the sharpest at the top of the superconducting dome (highest $T_c$), suggesting that at this point superconductivity is optimum regardless of whether or not there exists a magnetic quantum critical point (QCP). 

\begin{figure}
	\centering
	\scalebox{0.4}{\includegraphics{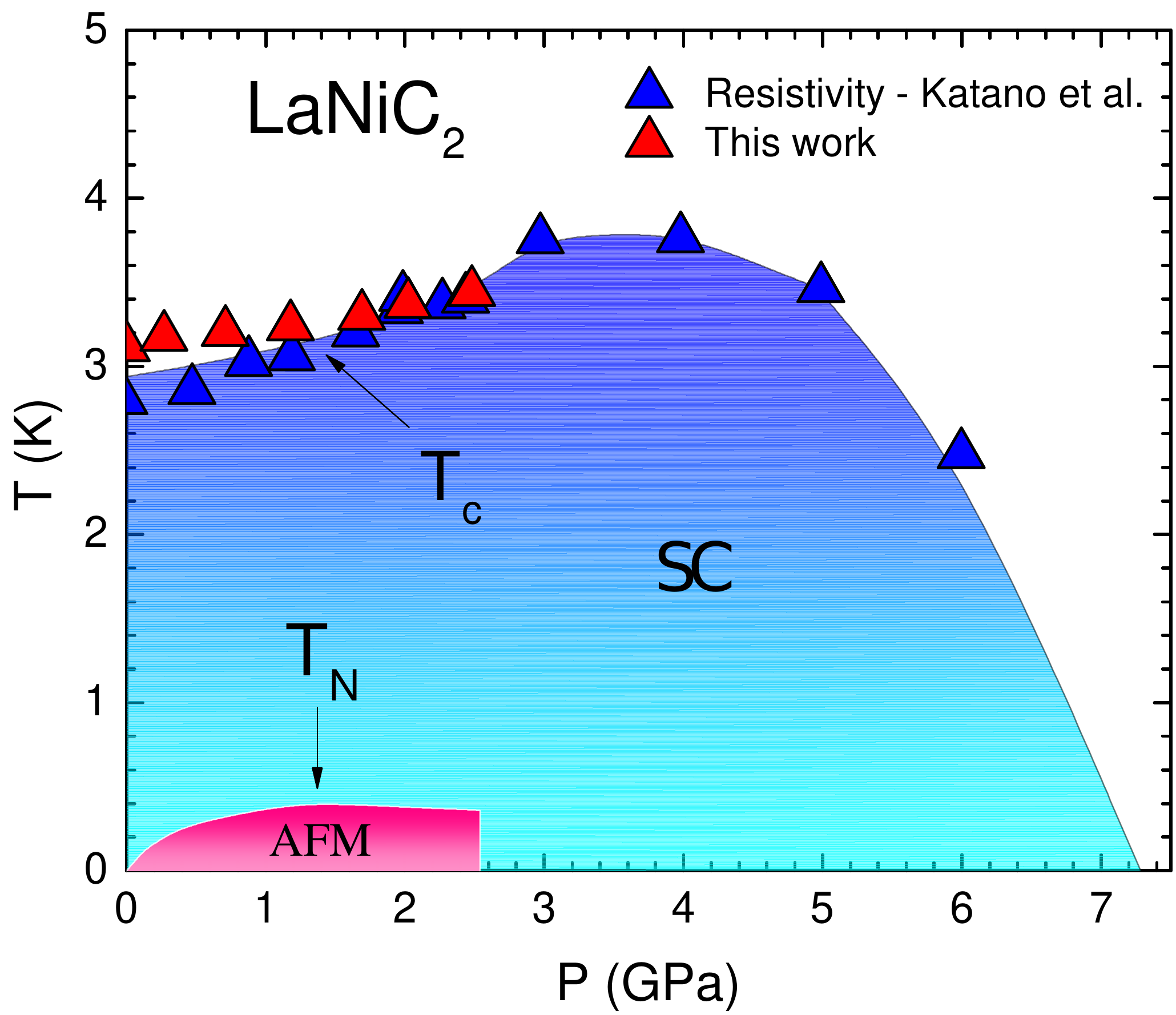}}
	\caption{\label{fig:Phase diagram}{(Color online) Phase diagram of LaNiC$_2$, including an antiferromagnetic phase with a quantum critical point inside the superconducting dome.}}
\end{figure}

Figure~\ref{fig:Pressure Sample A-1}(b) displays the low-temperature in-plane penetration depth data under pressure. For any of the applied pressure as temperature is lowered from 1.4 K $\Delta \lambda_{\parallel}(T)$ increases by as much as 0.16\% of the total signal, peaks in the range $0.2 - 0.5$ K, and then drops steeply. This pair-breaking behavior followed by a diamagnetic drop strongly indicates the emergence of a magnetic order within a superconducting phase.\cite{Jacobs1995,Schottl1999,Chia2001,Chia2005,Landaeta2017} The results provide the first-ever evidence of a magnetic phase in LaNiC$_2$, which is part of a large family $R$NiC$_2$ whose members $R$=rare-earth show antiferromagnetism at ambient pressure.\cite{Kotsanidis1989,Schafer1992} For comparison, we show in the inset of this figure pressure data of a cadmium sample taken in the same experimental system. It is worth mentioning the absence of magnetic impurities in our sample, as no upturn is observed at ambient pressure and no sign of their presence was observed in an electron paramagnetic resonance spectrum (for comparison see Ref.~\onlinecite{Bonalde2011}). Line-node-induced Andreev bound states are also discarded. Furthermore, upturns due to impurities and bound states just increase without turning down at all as $T$ goes down. A new $T-P$ phase diagram, showing the coexistence of magnetic and superconducting phases, is sketched in Fig.~\ref{fig:Phase diagram}. A magnetic QCP develops around ambient pressure, with a magnetic order persisting up to the highest applied pressure of 2.48 GPa. This QCP may have profound implications on the driving mechanism of superconductivity in LaNiC$_2$.

The appearance of a magnetic phase in LaNiC$_2$ is as highly relevant as intriguing. Electronic band-structure calculations indicate that at ambient pressure most of $3d$-Ni orbitals are far from the Fermi level,\cite{Hase2009,Subedi2009} which is all consistent with the nonmagnetic character found in experiments.\cite{Iwamoto1998,Pecharsky1998} To gain insight into the appearance of the magnetic phase we carried out first-principles numerical calculations of the electronic density of states (DOS) and band structure in the pressure range 0-2 GPa. We find no evidence in the partial DOS of overlapping states near the Fermi level, which could be indicative of hybridizations. Neither we observe any significant variations in the band structure that could suggest a rearrange of the electronic states. All this is consistent with recent calculations for higher pressures up to 15 GPa.\cite{Wiendlocha2016} Spin-polarized electronic structure calculations under external pressure may need to be completed to consistently analyze magnetism in this compound.

\section{Final remarks}

Finally, we discuss the very pertinent point of the superconducting mechanism in LaNiC$_2$. Conventional superconductivity is firmly known to be mediated by electron-phonon interactions, whereas unconventional behavior is believed to be caused by the interaction of local magnetic moments and conduction electrons (i.e., be magnetically mediated). Point nodes in the energy gap and TRS breaking make superconductivity definitely unconventional in LaNiC$_2$. The magnetic phase and the QCP found in the present work inside the superconducting state fulfill the requirement by the common believe that magnetic instabilities lead to unconventional superconductivity. So the superconductivity picture of LaNiC$_2$ is quite consistent with that of most strongly correlated electron systems, such as cuprates, iron-pnictides, organics, and heavy fermions. On the other hand, there is the idea that in LaNiC$_2$ other contributions to the driving mechanism are in play or that another high-pressure phase that competes with superconductivity may possibly exist.\cite{Katano2014,Wiendlocha2016} This high-pressure phase could very well be the magnetic one found in the present work.

One expects that in centrosymmetric compounds an unconventional superconductivity in a $p$-wave triplet state would be favored by ferromagnetic instabilities,\cite{Rice1995,Mazin1997} whereas a $d$-wave singlet state would be induced by antiferromagnetic fluctuations.\cite{Mazin1999} In noncentrosymmetric LaNiC$_2$ the situation is a bit more complicated. The point nodes along the $a$-axis in the superconducting energy gap are consistent with the $A_2$ state which is a mixture of singlet and triplet states with no dominant component. So either anti- or ferromagnetism may cause pair-breaking in the superconducting phase of LaNiC$_2$. 

\section{Conclusions}

In summary, we reported measurements of the magnetic penetration depth of a high-purity single crystal of noncentrosymmetric LaNiC$_2$. The low-temperature responses suggest the presence of point nodes in the superconducting energy gap. At finite pressure an antiferromagnetic order appears inside the superconducting phase below 0.5 K, revealing the presence of a magnetic quantum critical point at ambient pressure. Our results enhance the idea that unconventional superconductivity appears close to magnetic instabilities.

\begin{acknowledgments}
We thank T. Komatsubara for assistance in the single-crystal growth and Ney Luiggi for support in the calculations of the electronic structure and density of states. We acknowledge support from the Venezuelan Institute for Scientific Research (IVIC) grant No. 441 and the JSPS KAKENHI grant No. 15K05156.
\end{acknowledgments}


\end{document}